\documentclass[onecolumn,floats,prb,amsmath,amssymb,superscriptaddress, preprint]{revtex4-1}

\usepackage{graphicx}% Include figure files
\usepackage{bm}% bold math
\usepackage{color}
\usepackage{hyperref}
\usepackage{ulem}
\usepackage{amsmath}
\usepackage{amssymb}
\usepackage{mathtools}
\usepackage{mathrsfs}

 \usepackage{here}
 \usepackage[dvipsnames]{xcolor}

\newcommand{\fex}{f_{\rm ex}}
\newcommand{\mr}{\mathbf r}

\newcommand{\bfr}{\boldsymbol{\mathbf{R}}}

\newcommand{\bfR}{\boldsymbol{\mathbf{r}}}
\newcommand{\bfrn}{\boldsymbol{\mathbf{R}}^M}
\newcommand{\bfqm}{\boldsymbol{\mathbf{q}}^N}
\newcommand{\bfRm}{\boldsymbol{\mathbf{r}}^N}

\begin{document}

\title{Structure and position-dependent properties of inhomogeneous suspensions of responsive colloids}

\author{Yi-Chen Lin}
\affiliation{Applied Theoretical Physics-Computational Physics, Physikalisches Institut, Albert-Ludwigs-Universit\"at Freiburg, D-79104 Freiburg, Germany}
\author{Benjamin Rotenberg}
\affiliation{Sorbonne Universit\'e, CNRS, UMR 8234 PHENIX, 75005 Paris, France}
\author{Joachim Dzubiella}
\affiliation{Applied Theoretical Physics-Computational Physics, Physikalisches Institut, Albert-Ludwigs-Universit\"at Freiburg, D-79104 Freiburg, Germany}
%\affiliation{Cluster of Excellence livMatS @ FIT - Freiburg Center for Interactive Materials and Bioinspired Technologies, Albert-Ludwigs-Universit\"at Freiburg, D-79110 Freiburg, Germany}
%\affiliation{Research Group for Simulations of Energy Materials, Helmholtz-Zentrum Berlin, D-14109 Berlin, Germany}

\date{\today}

\begin{abstract}

Responsive particles, such as biomacromolecules or hydrogels, display a broad and polymodal distribution of conformations and have thus the ability to change their properties (e.g, size, shape, charge density, etc.) substantially in response to external fields or to their local environment (e.g., mediated by cosolutes or pH).  Here, we discuss the basic statistical mechanics for a model of responsive colloids (RCs) by introducing an additional 'property' degree of freedom as a collective variable in a formal coarse-graining procedure.  The latter leads to an additional one-body term in the coarse-grained (CG) free energy, defining a single-particle property distribution for an individual polydisperse RC. We argue that in the equilibrium thermodynamic limit such a CG system of RCs behaves like a conventional polydisperse system of non-responsive particles.  We then illustrate the action of external fields, which impose local (position-dependent) property distributions leading to non-trivial effects on the spatial one-body property and density profiles, even for an ideal (non-interacting) gas of RCs. We finally apply density functional theory in the local density approximation (LDA-DFT) to discuss the effects of particle interactions for specific examples of i) a suspension of RCs in an external field linear in both position and property, ii) a suspension of RCs with highly localized properties (sizes) confined between two walls, and iii) a two-component suspension where an inhomogeneously distributed (non-responsive) cosolute component, as found, e.g., in the studies of osmolyte- or salt-induced collapse/swelling transitions of thermosensitive polymers, modifies the local properties and density of the RC liquid.  

\end{abstract}

\maketitle
%\pacs{}

\section{Introduction}

The last few years have witnessed a growing interest in the fundamental and applied study of responsive particles and colloids (RCs), \cite{Stuart2010,Motornov,adaptive,shape,shape2,capsules,Richtering, Schurtenberger} and related responsive materials.~\cite{Stuart2010,origami, actuator, Heuser, Andreas}  Responsiveness in a soft material or liquid is a feature displayed, for example, by solvated polymers that have a critical solution temperature (CST) at which they sharply switch between two different physicochemical states. The switch can be induced by local stimuli such as temperature, pH, or the (osmotic) pressure of a cosolute,\cite{Stuart2010, Kawasaki1997, Kawasaki2000, Sasaki1999} i.e., in general some external field or an environment which modifies the interactions between the polymer monomers, resulting effectively in a shift of the local solvent quality. The stimuli-responsiveness can be harvested for tailoring functionality in applications, for instance, actuators, soft sensors, triggered drug release by nano-carriers, \cite{Stuart2010} or selective catalysis in polymeric nanoreactor particles.~\cite{Rafa1,msde2020modeling} The properties of a RC, for example, made from a thermosensitive polymer network (hydrogel), changes substantially at the CST: sharp size changes by a factor of two or three and accompanying jumps in the polymer density and stiffness of more than one order of magnitude are not untypical.~\cite{Richtering,Schurtenberger} 

If the responsiveness can be tamed and controlled by a sophisticated synthesis,  then a highly local (position-dependent) and specific response to the environment with respect to the function of the material could also be achievable. This leads to so-called 'smart' or programmable functional colloids with some pre-defined interactivity with the local environment.~\cite{Stuart2010, Heuser, Andreas} An example could be directed drug release by responsive colloidal carriers where the drug is liberated only at desired places (hot spots) over a very well-defined period of time.~\cite{drug} Another possible example would be a colloidal dispersion of catalytic nanoreactor RCs that speeds up or slows down a chemical reaction depending on its local environment,\cite{Rafa1,msde2020modeling} triggered by a threshold fluctuation of some chemical species in solution nearby. Highly localized and programmable response with respect to local fields and environment are thus key for the design of next-generation soft functional colloids dispersed in a suspension. 

%Activity may come in when the particles can produce energy and self-sustain some active motion or transformations, e.g., powered by an internal or surface chemical reaction as in biological cells. In the recent years the scientific field of active colloids, for instance, as model systems for living macromolecules, such as bacteria or moving cells, or to develop artificial functional swimmers has exploded.~\cite{Hartmut} The focus in these studies was exclusively put on active translational motion, i.e., some realization of self-propelling action. However, activity can in principle also mean that properties of the particle can actively change, e.g., the size, shape,  or conformational properties as well as the density or elasticity of a particle.~\cite{shape,origami} Depending on the internal energy generating mechanism this can be stochastic or oscillatory, or in response to an external stimulus, which brings us back to responsiveness and adaptivity, but now coupled to activity. 
 
Typical examples of soft and functional colloidal RCs, which will be focus of this work, are hydrogel or block-copolymer particles synthesized with stimuli-responsive polymers,~\cite{Stuart2010,Richtering,shape} and biomolecular (or bio-inspired) polymeric particle assemblies from DNA, peptides, and proteins.~\cite{protein,IDP_switching,DNA} Many properties of such a RC, for example the macromolecular conformation,\cite{protein,RC,IDP_switching,chiwu} size,\cite{matthias,denton:softmatter2016,Schurtenberger} shape,\cite{denton:JCP2014,denton:JCP2016,  denton:softmatter2015,shape,shape2,Schurtenberger,hartmut_pnipam} charge density,\cite{denton:softmatter2018,Schurtenberger} or dipole,\cite{Berne} etc., are thus responsive and highly fluctuating quantities. In other words, a single RC is intrinsically polydisperse, and any observable property $\sigma$ of a given RC follows a probability distribution, $p(\sigma)$.  The property $\sigma$ formally represents a coarse-grained, stochastic variable that results from integrating out the underlying microscopic (internal) degrees of freedom of the RC in some reference environment, e.g., a single hydrogel particle or protein in clear water. A simple and typical example for $\sigma$ is the radius of gyration of a polymer\cite{Fixman1962,kremer,matthias} or hydrogel.~\cite{Richtering, Scotti, Winkler, Emanuela,Schurtenberger}  In the realm of protein folding, $\psi(\sigma) = -k_BT \ln p(\sigma)$, typically is usually called the free energy landscape~\cite{protein,RC} for folding, where $\sigma$ is some other meaningful collective variable characterizing the structure, e.g., the protein fraction of native contacts or the root-mean-square deviation from the native fold. A functional response of such a particle to an inhomogeneous field or interacting environment means that the single-particle property distribution, $p(\sigma)$, will be modified and feed back to the spatial structure of the whole suspension via particle-particle interactions in some non-trivial and local (position-dependent) way. Indeed, reversible aggregation, clustering, phase-separation in suspensions of RCs induced and modified by stimuli have been demonstrated experimentally.~\cite{Motornov,Schurtenberger}

In standard theoretical studies of the structure of soft colloidal liquids the variable characterizing a property is not explicitly  resolved, i.e., all microscopic degrees of freedom are only implicitly contained in the effective pair potential between the colloids,~\cite{likos, SoftColloid} and the conventional position-dependent response to an external field.~\cite{hartmut_external} However, as argued above, the knowledge of the local property distribution and how it responds to fields and the interacting environment, including the interaction between RCs themselves, is the key to understanding functionality and structuring  of RC suspensions, and any modeling effort in this direction must resolve it.  Notable exceptions in the literature that included a property response are the works by Denton and Schmidt on colloid-polymer mixtures with compressible polymers~\cite{matthias} and subsequent works by Denton {\it et al.} on penetrable and shape-fluctuating polymers and compressible hydrogels.~\cite{denton:JCP2014, denton:JCP2016,denton:softmatter2015,denton:softmatter2016} In these works, the polymeric size and/or shape was considered as a specific property and its distribution and bulk response to the surrounding cosolute (hard colloids) was explicitly taken into account by an additional energetic one-body term in the Hamiltonian, either in a density functional theory (DFT) framework~\cite{matthias}  or in Monte-Carlo (MC) simulations.~\cite{denton:JCP2014, denton:JCP2016,denton:softmatter2015,denton:softmatter2016} However, no general framework for the explicit resolution of properties of RCs and for studying the RC liquid structure and property response under the action of external fields or inhomogeneous environments has been introduced yet.

The objectives of the present work are i), to derive the general basic statistical mechanics premises of a coarse-grained colloidal model of fluids of RCs, where in addition to the effective pair potential the single-particle polydispersity is explicitly considered as a coarse-grained random variable, and ii), to study how the latter affects the liquid structure in the presence of modulating fields and interacting environments.  At this point, it is important to emphasize that a system of RCs, which {\it individually} are responsive and polydisperse,  differs physically from the conventional picture of polydisperse systems where each individual particle property is fixed according to a polydisperse ensemble distribution. The latter was treated extensively in literature  but most prominently only for hard sphere colloids with ensemble distributions of the hard sphere size following typically a simple Gaussian polydisperse distribution.\cite{Salacuse1,Salacuse2, Briano1984, Barrat1986, Kofke1988, bartlett, Poon1998, Warren, IgnacioPRL2000,  Sollich2002} Under what circumstances RCs and conventional polydisperse systems behave the same or follow the same statistical rules is {\it a priori unknown.} We will actually demonstrate that in equilibrium in the grand canonical ensemble (more precisely, in the thermodynamic  limit, TDL) such a model system of RCs is equivalent to the statistical mechanics formulations of conventional polydisperse systems of non-responsive colloids (with fixed property per particle).   Hence, the resulting statistical mechanics  we describe in this work is in some limits already known, however, it provides interesting new perspectives and future prospects of polydispersity from the viewpoint of soft and functional responsive materials. 

In particular, we study some minimalistic examples to  illustrate the leading effects of responsiveness on the liquid structure.  For instance, a key consequence of the responsiveness and the intrinsic polydispersity of the RCs is that their property distributions and their means and moments are position-dependent in external fields and in inhomogeneous co-solvent environments. This is highly relevant for applications where a local property is needed and shall be selected by a field or environment for the desired function. Local size segregation in external fields, for instance, has already been observed for conventional polydisperse hard spheres confined between hard walls using DFT.~\cite{IgnacioPRL2000}  However, position-dependent colloidal properties resolved in space have not been discussed in detail before. We demonstrate for the first time that such a property localization is generic even for an ideal gas of RCs in external fields.  Moreover, the effective pair potential between the RCs is property-dependent, i.e., its action depends on the local property distribution in absolute space and not just on the relative particle pair distance. Effects of this we study using a simple (local density approximation, LDA) DFT for examples of interacting RCs in linear external fields.  We also present LDA results on the structure of a system of highly localized properties (sizes) confined between walls, not studied before in that limit. Hence, our work provides, from a new perspective, a deeper insight into the effects of external fields and pair interactions on local polydisperse properties, and how they affect the local density response. We finally provide an outlook on future research directions in the field, in particular, regarding observables or circumstances for which the resulting behavior of RCs will differ qualitatively from conventional polydisperse systems.

\section{Statistical Mechanics of Liquids of Responsive Colloids}

\subsection{Microscopic partition sum and coarse-graining}

We start by formally coarse-graining a one-component system of a number $N$ of RCs with the aim to define a coarse-grained property distribution function as well as the corresponding one- and two-body approximations of the free energy of the coarse-grained (CG) system. For this, we consider a set $\bfrn$ of microscopic coordinates, with $M = mN$ for $m$ monomers per RC (for simplicity we assume all monomers are identical). We want to introduce  a coarse-grained description consisting of collective variables for the $N$ RCs described by their positions of their centers of mass $\bfR_i = (1/m)\sum_{\alpha=1}^m \mathbf{R}_{i\alpha}$, $i=1..N$, as in conventional polymer coarse-graining,~\cite{HansenMcDonaldbook} and also an additional CG degree of freedom $\sigma_i$.  The latter in the following we term a {\it particle property} and it could be identified, for example, with particle size, shape, charge density, etc. In general, we could define a set of orthogonal (independent) properties, or a property vector, but for the sake of simplicity we focus only on a single CG property in this work.  The CG description is then specified by the collective variables $\bfqm=(\bfRm,\sigma^N) = (\mr_1,..,\mr_N,\sigma_1,..,\sigma_N)$. The canonical partition function of the RCs with the microscopic interaction Hamiltonian for the monomers, $U( \bfrn )$, and without an external field is
\begin{align}
\label{eq:Z}
\mathcal{Q} &=
\frac{ 1 }{\Lambda_m^{3M} M!}
\int {\rm d}\bfrn
e^{-\beta U( \bfrn ) },
\end{align}
where $\Lambda_m$ is the thermal wavelength of the monomers. We introduce the collective variables in the integral
as~\cite{HansenMcDonaldbook}
\begin{align}
\mathcal{Q} &=
\frac{ 1 }{\Lambda_m^{3M} M!}
\int {\rm d}\bfrn
\int {\rm d}\bfRm
\int {\rm d}\sigma^N
\prod_{i=1}^{N} \delta( \bfR_i(\bfr^M)  - \bfR_i )  
\prod_{i=1}^{N} \delta( \sigma_i(\bfr^M)  - \sigma_i )  
e^{-\beta U( \bfrn ) },
\end{align}
where we separate the microscopic configurations
corresponding to prescribed values of the collective variables,
and integrate over all possible values of the latter.
We then rewrite as
\begin{align}
\label{eq:Zcg}
\mathcal{Q} &=
\frac{ p_0^N }{\Lambda_{cg}^{3N} N!}
\int {\rm d}\bfRm
\int {\rm d}\sigma^N
e^{-\beta \mathcal{F}( \bfRm,\sigma^N ) }, 
\end{align}
where we have introduced the probability 
\begin{align}
\label{eq:Fdef}
e^{-\beta \mathcal{F}( \bfRm, \sigma^N ) }
&=
\frac{\Lambda_{cg}^{3N} N!}{p_0^N \Lambda_m^{3M} M!}
\int {\rm d}\bfrn
\prod_{i=1}^{N} \delta( \bfR_i(\bfr^M)  - \bfR_i )  
\prod_{i=1}^{N} \delta( \sigma_i(\bfr^M)  - \sigma_i )  
e^{-\beta U( \bfrn ) }
\end{align}
according to the free energy $\mathcal F$ of the CG state $\bfqm=(\bfRm,\sigma^N)$, and $\Lambda_{cg}$  the de Broglie wavelength corresponding to the total mass of each CG responsive particle (which is irrelevant for the calculation of mean observables). To make the partition sum $\mathcal{Q}$ properly dimensionless we also introduced $p_0$, which plays the role of an inverse microscopic 'unit' property that is the basis for counting the number of property states. It will serve later in the applications simply as as a normalizing prefactor like the de Broglie wavelength (being the unit for the 1D Euclidean space) in conventional treatments. 

In principle we should consider the joint distribution
of all collective variables $\bfqm=(\bfRm,\sigma^N)$ for applications,
but we can start making simplifying assumptions which
will certainly be reasonable in the low density limit,
such as assuming that effective interactions  -- as defined
by the free energy $\mathcal{F}( \bfRm, \sigma^N)$ --
can be expressed as a sum of one- and two-body contributions,~\cite{likos,HansenMcDonaldbook,bolhuis2001} through
\begin{eqnarray}
\mathcal{F} ( \bfRm, \sigma^N ) \simeq  F_0(N/V) + \sum_i^N \psi(\sigma_i) + \frac{1}{2} \sum_{i\neq j}^N \phi(\mr_i,\mr_j; \sigma_i.\sigma_j), 
\label{eq:pp} 
\end{eqnarray}
This equation defines the property (free) energy landscape $\psi(\sigma)$ of a single, isolated RC without any external field through the probability distribution
\begin{eqnarray}
p(\sigma) = p_0 \exp[-\beta \psi(\sigma)] = \frac{\Lambda_{cg}^{3}}{\Lambda^{3m}_m m!}\int {\rm d}\bfr^m \delta( \sigma(\bfr^m)  - \sigma)  e^{-\beta U( {\bf R}^m ) },
\label{Eq:single} 
\end{eqnarray}
The term $\sum_i^N \psi(\sigma_i)$  in eq.~(\ref{eq:pp}) represents a one-body term as an explicit function of the properties realized in a micro-configuration of the $N$ RCs. The volume term~\cite{bolhuis2001} in eq.~(\ref{eq:pp}), $F_0(N/V)$, is independent of the configuration $\{\mr_i,\sigma_i\}$ and includes the kinetic terms of the monomers.  The last term in eq.~(\ref{eq:pp}) defines the pair potential $\phi$ of the system, depending not only on the positions of the two particles but also their instantaneous properties. Note that this definition is different as in conventional studies of coarse-grained potentials because the one-body term carries explicitly the energetic changes associated with property changes in the interacting system, which are usually integrated out. We will discuss this fact in more detail in section~\ref{2body}.

We now consider the action of an external field. In the case of ideal (non-interacting) RCs,
which only have intra-molecular interactions and coupling to the field, we can write the potential energy
as $U(\bfr^M) \equiv U^{\rm id}(\bfr^M)=\sum_{i=1}^N\sum_{\alpha=1}^m \left[ u_{\rm intra}(\{\bfr_{i\alpha}\}) +
\phi_{\rm ext}(\{\bfr_{i\alpha}\}) \right]$, i.e., the external field $\phi_{\rm ext}$ acts on each monomer. 
Integrating out the microscopic degrees of freedom for each RC
then defines the joint distribution function $\rho_{\rm id}(\bfR,\sigma)$ 
which in the presence of the external potential depends on the position of the center of mass {\it and} the property.  If we assume that the external potential varies only over large distances compared to the typical size of an RC, we can
change coordinates relative to the centers of mass, 
$\delta \bfr_{i\alpha}=\bfr_{i\alpha}-\bfR_i$, and coarse-grain
$\phi_{\rm ext}(\bfr_{i\alpha})$ into
\begin{eqnarray}
\phi_{\rm ext}(\bfR_i, \sigma_i) = -k_BT \ln \langle \exp(-\beta \phi_{\rm ext}(\bfr_{i\alpha}))\rangle_{\alpha}
\end{eqnarray}
where the average is taken over all monomers $\alpha$ of a single RC at fixed $\bfR_i$ and $\sigma_i$. Hence, we can rewrite the free energy for ideal RCs in an external field approximately as
\begin{align}
e^{-\beta \mathcal{F_{\rm id}}( \bfRm,\sigma^N ) }
&=
\frac{\Lambda_{cg}^{3N} N!}{\Lambda^{3M} M!}
\prod_{i=1}^{N} 
e^{-\beta \phi_{\rm ext}(\bfR_i, \sigma_i) }
\left(\prod_{\alpha=1}^m 
\int {\rm d}\delta\bfr_\alpha\right)
\delta( \sigma_i(\{\delta\bfr_{i\alpha}\})  - \sigma_i )  
e^{-\beta u_{\rm intra}(\{\delta\bfr_{i\alpha}\}|\bfR_i)}
\end{align}
%Up to some normalization constant, the multiple integrals over the internal
%degrees of freedom define a term per particle which corresponds to
%$\rho_{\rm id}(\sigma|\bfR)$. The latter finally reduces to $p(\sigma)$ if one choses the reference state
%(with respect to which $\phi_{\rm ext}$ is defined). 
As a result, for an interacting system of RCs in an external field we can employ the following approximation for the total coarse-grained free energy for a given macro-configuration $\bfqm=(\bfRm,\sigma^N)$
\begin{eqnarray}
\mathcal{F} ( \bfRm, \sigma^N ) \simeq  F_0(N/V)+\sum_i^N [\psi(\sigma_i) + \phi_{\rm ext}(\mr_i,\sigma_i)] + \frac{1}{2} \sum_{i\neq j}^N \phi(\mr_i,\mr_j; \sigma_i,\sigma_j). 
\label{CGHam}
\end{eqnarray}
The last term defines a pair potential which is explicitly property-dependent.  The one-body property landscape $\psi(\sigma) = -k_BT\ln [p(\sigma)/p_0]$ appeared already in literature in a 'responsive' context in CG polydisperse Hamiltonians to specifically describe compressible polymer chains in classical DFT.~\cite{matthias}  Related terms were also presented in CG Hamiltonians for fluctuating blobs in polymer chains,~\cite{kremer} shape-fluctuating polymers,\cite{denton:JCP2014,denton:JCP2016} or for deformable hydrogels.\cite{denton:softmatter2016}  We note that the property $\sigma$ can also represent a particle orientation (i.e., an angle w.r.t. some reference direction in 2D). In that well-studied limit, regarding the orientation as a polydisperse attribute, the singe-particle distribution $p(\sigma)$ would be simply a continuous uniform (or 'rectangular') distribution,~\cite{HansenMcDonaldbook} as all orientations for a single particle would be equally likely in some interval, for example $\sigma\in[0:2\pi]$ in 2D. 

\subsection{Partition functions of coarse-grained RC liquids}

The single-particle property distribution, $p(\sigma)$, is in general not necessarily Gaussian nor narrow for RCs  and can feature even multiple metastable states, such as well known in protein folding.~\cite{protein,RC} Another example are responsive hydrogels which display a distinct two-state behavior of swollen and collapsed configurations close to their critical solution temperature.\cite{Stuart2010,Richtering} 
As argued above, for such a distribution $p(\sigma)$ we can assign a property potential energy function 
\begin{eqnarray}
\psi (\sigma)= -k_BT\ln[p(\sigma)/p_0],
\end{eqnarray}
which describes the (free) energy of being in a property state $\sigma$. We chose $p_0$ conveniently as
\begin{eqnarray}
p_0^{-1} = \int_{-\infty}^\infty {\rm d}\sigma \exp[-\beta {\psi (\sigma)]}, 
\label{norm}
\end{eqnarray}
such that the probability distribution is normalized, $\int_{-\infty}^\infty {\rm d}\sigma p(\sigma)=1$. The mean property of such a distribution for a single RC is then
\begin{eqnarray}
\bar \sigma = \int_{-\infty}^\infty \sigma p(\sigma) {\rm d}\sigma. 
\end{eqnarray}
Higher moments, for example to calculate the variance of the property, can be defined in the usual way, i.e., $\overline {\sigma^m} = \int \sigma^m p(\sigma)d\sigma$.  If we consider the limit
\begin{eqnarray}
p(\sigma) = \delta(\sigma-\sigma_0), 
\end{eqnarray}
using the Dirac $\delta$-function, then the RCs have only a single property $\sigma_0$ as in a conventional one-component system. 
%
%The canonical partition sum of a single RC freely moving in a volume $V$ where the property is subject to thermal fluctuations reads
%\begin{eqnarray}
%Z_1 = \frac{p_0}{\Lambda^3}\int_V {\rm d}{\mathbf r} \int_{-\infty}^\infty {\rm d}\sigma \exp[-\beta {\psi (\sigma)]} = \frac{V}{\Lambda^3}Z_\sigma, 
%\label{Z1}
%\end{eqnarray}
%where we use $\Lambda$ now as the de Broglie wavelength of one RP, and we need to integrate over all possible realizations of $\sigma$.  If we consider the limit
%\begin{eqnarray}
%p(\sigma) = \delta(\sigma-\sigma_0), 
%\end{eqnarray}
%using the Dirac $\delta$-function, then we have only one single property $\sigma_0$ of the RC and $Z_\sigma=1$. 
%Eq.~(\ref{Z1}) can be easily extended to 

Given the coarse-gained free energy eq.~(\ref{CGHam}), the canonical partition sum for $N$ interacting RCs in an external field is given by the weighted integral over all states
\begin{eqnarray}
Z_N = \frac{p_0^N}{\Lambda^{3N} N!}\prod_i^N \left\{ \int_V {\rm d}{\mathbf r_i} \int_{-\infty}^\infty {\rm d}\sigma_i \right\} \exp[-\beta{\cal H}_N(\{\mathbf r, \sigma\})-\sum_i^N \beta \psi (\sigma_i)].
\label{ZN}
\end{eqnarray}
As demonstrated in the previous section, the position-dependent Hamiltonian ${\cal H}_N$ (within a constant) can be approximated by a sum of the potential energy due to pair interactions and external fields through
\begin{eqnarray}
{\cal H}_N(\{\mathbf r, \sigma\}) = \frac{1}{2} \sum_{i\neq j}^N \phi(\mr_i,\mr_j; \sigma_i,\sigma_j) + \sum_i^N \phi_{\rm ext}(\mathbf r_i, \sigma_i). 
\end{eqnarray}
The pair potential $\phi(\mr_i,\mr_j; \sigma_i,\sigma_j)$ depends on both the positions and the  properties of the two interacting RCs.  The $\phi_{\rm ext}(\mathbf r_i, \sigma_i)$ is the conventional external field which as usual is a function of particle position and its property (such as a charge or dipole in an external electrostatic field). Thus, the set $\{\mathbf r_i, \sigma_i\}$ denotes the full configuration, i.e., the RC position vectors and properties. 

For an ideal gas of RCs and no external field (${\cal H}_N\equiv 0$) the Helmholtz free energy is $F_{\rm id} = Nk_BT[\ln (\rho\Lambda^3)-1]$ with constant number density $\rho=N/V$, and the corresponding total chemical potential $\mu_{\rm id} = k_BT\ln(\rho\Lambda^3)$. We recognize that the property energy landscape $\psi$ in eq.~(\ref{ZN}) formally plays the role of an external field.~\cite{HansenMcDonaldbook,matthias} Hence, the total chemical potential of an ideal gas of RCs can also be expressed by $\mu_{\rm id} = k_BT\ln[\rho_{\rm id}(\sigma)\Lambda^3/p_0)] + \psi(\sigma)$, introducing the ideal gas property density  $\rho_{\rm id}(\sigma)$ (with units per volume per property and is homogeneous in position space). The property density distribution for the ideal RC gas thus follows a law analogous to the barometric height law $\rho_{\rm id}(\sigma) = \rho p(\sigma) = \frac{p_0 e^{\beta \mu_{\rm id}}}{\Lambda^3}\exp(-\beta \psi(\sigma))$. Pair potentials and the conventional position-dependent external field will modify the ideal distribution $p(\sigma)$ to an emerging distribution $N(\sigma)$ as demonstrated in the next sections. 

%
%We recognize that the second term on the right hand side of eq.~(\ref{Eq:Fid}) plays the role of an additional part of the chemical potential in the ideal gas limit where only one-body terms contribute. Hence, the chemical potential of an ideal gas of RCs  is 
%\begin{eqnarray}
%\mu_{\rm id} =  \mu_0 - k_BT \ln Z_\sigma = \mu_0 - k_BT\ln p_0\int {\rm d}\sigma \exp[-\beta \psi(\sigma)], 
%\label{chempot}
%\end{eqnarray}
%where $\mu_{\rm id} = k_BT\ln(\rho\Lambda^3)$ with $\rho = N/V$; $\mu_0$ is the purely translational ideal gas reference contribution for which we can define a reference density $\rho_0 = \exp(-\beta \mu_0)/\Lambda^3$.  It is in fact clear from partition sum eq.~(\ref{ZN}) that the change in the chemical potential for shifting a property is $\mu_{\rm id}(\sigma) \propto -\psi(\sigma)$, that is, the one-particle landscape $\psi$ plays the role of a (negative) chemical potential contribution for a particle of property $\sigma$. 

We can finally define the corresponding grand partition function by introducing the total chemical potential $\mu$ of RCs in the grand canonical ensemble, via   
\begin{eqnarray}
{\cal Z} = \sum_{N=0}^\infty  \frac{q^N}{N!}\prod_i^N \left\{ \int_V {\rm d}{\mathbf r}_i \int_{-\infty}^\infty {\rm d}\sigma_i \right\} \exp[-\beta {\cal H}_N(\{\mathbf r, \sigma\})-\sum_i^N\beta \psi(\sigma_i)]
\label{GC}
\end{eqnarray}
with $q =p_0\exp(\beta \mu)/\Lambda^3$ being the fugacity.  The grand potential follows from $\Omega = -k_BT\ln \cal Z$. For the RC ideal gas in the grand canonical ensemble $q=\rho = \langle N \rangle /V$, where the angular brackets denote the ensemble average. Given the analogy between the property potential and the external field, we can rewrite eq.~(\ref{GC}) in terms of the chemical potential of the RCs with property $\sigma$ in the reservoir, $\mu(\sigma) = \mu - \psi(\sigma)$,~\cite{HansenMcDonaldbook,matthias} which represents the contribution to the total chemical potential $\mu$ that is not explicitly dependent on $\psi$:  
\begin{eqnarray}
{\cal Z} = \sum_{N=0}^\infty  \frac{p_0^N}{\Lambda^{3N}N!}\prod_i^N \left\{ \int_V {\rm d}{\mathbf r}_i \int_{-\infty}^\infty {\rm d}\sigma_i \right\} \exp[-\beta {\cal H}_N(\{\mathbf r, \sigma\})+\sum_i^N\beta \mu(\sigma_i)]
\label{GC2}
\end{eqnarray}

% and is a functional of $\psi$, $\Omega[\psi]$. Considering the functional derivative of the latter with respect to $\psi$, it is straightforward to show that
%\begin{eqnarray}
%\frac{\delta \Omega[\psi]}{\delta\psi} = \left\langle \sum_i \delta(\sigma - \sigma_i)\right\rangle = N(\sigma), 
%\end{eqnarray}f
%which defines the emerging property distribution, $N(\sigma)$, resulting from the interactions according to the full Hamiltonian ${\cal H}_N$. (The bracket denotes the grand canonical ensemble average).  In the ideal gas limit and without external potential the grand potential takes the form $\Omega_{\rm id} = -k_BT\rho V$, and $N(\sigma) = \rho Vp(\sigma)$ with the mean number density $\rho = N/V = \exp(\beta\mu_{\rm id})/\Lambda^3$.  
%
%The mean particle number is $N = \int {\rm d}\sigma N(\sigma)$. 
 
It is instructive to compare eq.~(\ref{GC2}) to the grand partition function of a conventional polydisperse fluid with a continuous polydispersity distribution of a property $\sigma$ defined by a prescribed chemical potential distribution $\mu(\sigma)$ of the ensemble in the reservoir.~\cite{Briano1984, Kofke1988,Sollich2002} We find that the grand partition function for the latter case, exactly true in the thermodynamic limit (TDL), reads the same (within some arbitrary constant of the reference free energy) as eq.~(\ref{GC2}). This means that -- at least regarding the equilibrium structure in the TDL -- the statistical mechanics treatment of a polydisperse system of non-responsive particles (i.e, with the property $\sigma$ fixed for every individual particle) with an ensemble property distribution $N(\sigma)$ (often 'parent' distribution~\cite{IgnacioPRL2000,  Sollich2002}) and a pair potential $\phi$ is the same as the one of a system of identical (one-component) RCs with a distribution of properties $N(\sigma)$ per particle, interacting with the same pair potential $\phi$.  The analogy makes sense, however, if we recall the Monte-Carlo scheme of sampling equilibrium states,~\cite{Kofke1988,Sollich2002} where particles and their properties can be simply switched (by exchanges with the reservoir) to sample phase space according to the correct ensemble weights.  Salacuse in fact argued that random realizations of the size of a single hard sphere and a distribution of a polydisperse ensemble of various hard spheres (but individually fixed sizes) are the same in the TDL.~\cite{Salacuse2} As a consequence we can adopt the established knowledge about conventional polydisperse systems to the case of responsive colloids. We will see in the following, however, that the perspective and some interpretations of the results are different. Also, property changes are usually accompanied by changes in the potential energy, which has not been in detail discussed before, and they also have some implications for the definition of a consistent effective pair potential for interacting RCs.   

\subsection{The one- and two-particle number and property density distributions}

\subsubsection{One-body distribution functions}
The one-body property-density distribution function (PDDF) of the RCs can be defined formally as 
\begin{eqnarray}
\rho(\mathbf r,\sigma) = \left\langle \sum_{i=1}^N \delta(\mathbf r - \mathbf r_i)\delta(\sigma-\sigma_i)\right\rangle, 
\end{eqnarray}
where the brackets $\langle..\rangle$ denote the ensemble average according to the grand canonical partition function introduced above. The one-body number density distribution of RCs  follows as 
\begin{eqnarray}
\rho(\mr) = \int_{-\infty}^{\infty} {\rm d}\sigma \rho(\mathbf r, \sigma)
\end{eqnarray}
and the property distribution as
\begin{eqnarray}
N(\sigma) = \int_V {\rm d}\mr \rho(\mathbf r, \sigma).
\end{eqnarray}
From standard calculations using the definition of the ensemble averages, one can show that $N(\sigma){\rm d}\sigma = {\rm d}N(\sigma)$ denotes the number of particles with property $\sigma$ in the system. The total particle number is provided by $\langle N \rangle =\int {\rm d}\sigma N(\sigma)$. Moreover, for a homogeneous system in the absence of an external field, $\rho(\mr, \sigma) = N(\sigma)$/V. In conventional polydisperse {\it bulk} systems, $N(\sigma)$ is often called the "parent distribution" because it is the distribution in the reservoir of interacting particles.~\cite{IgnacioPRL2000,  Sollich2002} We  argue that it would be better to call $N(\sigma)$ the {\it emerging} property distribution, originating from a modification of the ideal distribution $p(\sigma)$ due to interactions and/or the action of an external field. In that sense, $p(\sigma)$ is more intrinsic and could also serve as a better definition of a parent.  Note that one should compare $p(\sigma)$ to $N(\sigma)/\langle N \rangle$ to be consistent with normalization.  For the case of the homogeneous ideal gas of RCs, $N_{\rm id}(\sigma)=\rho V p(\sigma)$, and thus $\rho_{\rm id}(\mr, \sigma) = \rho_{\rm id}(\sigma) = \rho p(\sigma)$.

 In the presence of an external field there exists a spatially varying mean property (and the corresponding higher moments), according to 
\begin{eqnarray}
\sigma(\mr) = \frac{1}{\rho(\mr)}\int_{-\infty}^\infty {\rm d}\sigma\, \sigma \rho(\mr,\sigma).
\label{sigr}
\end{eqnarray}
 The global mean property of a system of RCs consequently is 
\begin{eqnarray}
\bar \sigma &=&\frac{1}{N}\int_V {\rm d}\mr \int_{-\infty}^\infty {\rm d}\sigma \sigma \rho(\mr,\sigma)  \nonumber \\
&=& \frac{1}{N}\int_{-\infty}^\infty \sigma N(\sigma) {\rm d}\sigma = \frac{1}{N}\int_V {\rm d}\mr \rho(\mr)\sigma(\mr).
\end{eqnarray}

\subsubsection{Ideal RC distributions in an external field}

For an ideal (non-interacting) gas of RCs in an external field we find explicitly
\begin{eqnarray}
\rho_{\rm id}(\mr,\sigma) &=& \frac{\exp(\beta\mu_{\rm id})}{\Lambda^3}p(\sigma) \exp[-\beta \phi_{\rm ext}(\mr, \sigma)] \nonumber \\
&=& q\exp[-\beta \psi(\sigma) - \beta \phi_{\rm ext}(\mr, \sigma)]
\label{idgas}
\end{eqnarray}
and accordingly for the one-body density distribution
\begin{eqnarray}
\rho_{\rm id}(\mr) = q \int_{-\infty}^{\infty} {\rm d}\sigma\exp[-\beta \psi(\sigma) - \beta \phi_{\rm ext}(\mr, \sigma)],
\label{1b}
\end{eqnarray}
and for the property distribution
\begin{eqnarray}
N_{\rm id}(\sigma) = q\exp[-\beta \psi(\sigma)]\int_V {\rm d}\mr  \exp [- \beta \phi_{\rm ext}(\mr, \sigma)].
\end{eqnarray}

\begin{figure}[ht!]
   \centering
      \includegraphics[width=0.32\textwidth]{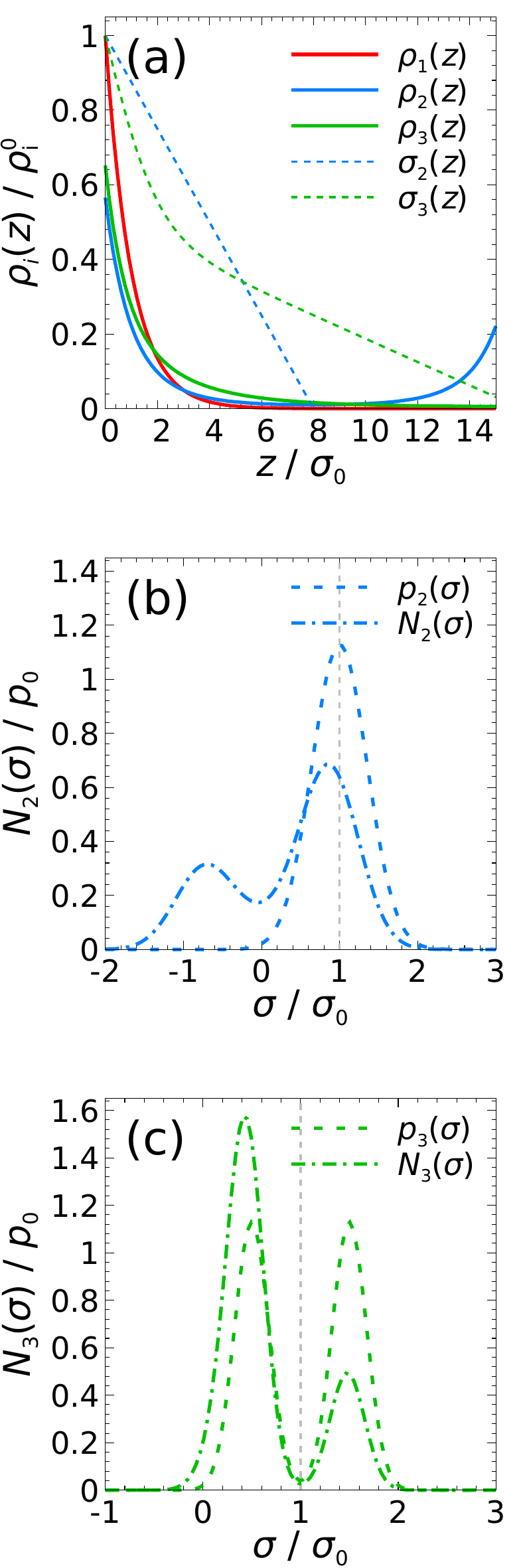}
   \caption{(a) Number density profiles $\rho_i(z)/\rho_i^0$ of an ideal gas of RCs (normalized such that $N$=const.) confined between walls at $z=0$ and $L=15$ in a linear external potential $\phi_{\rm ext} = A\sigma z$ with single-particle property distributions according to eqs.~(\ref{p1}) to (\ref{p3}). The external field parameter is $\beta A=1$. The dotted lines in (a) are the spatial property profiles $\sigma_i(z)$ using the same scale than the normalized density. 
   (b) The emerging property distribution $N_2(\sigma)$ for the Gaussian ideal gas distribution $p_2(\sigma)$, cf. eq.~(\ref{p2}),  in the external field. (c) The property distribution $N_3(\sigma)$ for the double Gaussian ideal gas distribution $p_3(\sigma)$, cf. eq.~(\ref{p3}), in the external field.}
   \label{fig:id}
\end{figure}

We discuss now a few pedagogical examples of density distributions of an ideal gas of RCs in an external field. For simplicity and illustration purposes we chose an external field linear in both position and property, according to 
\begin{eqnarray}
\phi_{\rm ext} = \sigma Az, 
\end{eqnarray}
where $A>0$ is a constant, and $z$ is the position in one cartesian direction, that is, we just focus on inhomogeneities in one spatial dimension. $Az$ plays the role of an external potential conjugate to property $\sigma$. External potentials linear in both property and its conjugated field are very common in physics, for example, a charge $q$ in a linear electrostatic potential $Ez$, or a dipole $d$ in a linear electric field $E'z$.  We ensure an equilibrium, zero flow situation  by confining the system between hard walls at $z=0$ and $L$. We compare the following normalized single-particle property distributions
\begin{eqnarray}
\label{p1}
p_1(\sigma) &=& \delta (\sigma - \sigma_0), \\  \label{p2}
p_2(\sigma) &=& \frac{1}{\sqrt{2\pi\tau^2}} \exp\left[-(\sigma-\sigma_0)^2 / (2\tau^2)\right], \;\;{\rm and} \\ \label{p3}
 p_3(\sigma) &=& \frac{1}{2\sqrt{2\pi\tau^2}}  \sum_{i=1,2}\exp[-(\sigma-\sigma_i)^2 / (2\tau^2)].
\end{eqnarray}
$p_1(\sigma)$ is simply the Dirac $\delta$-distribution and fixes the RC property to a single value $\sigma_0$; $p_2(\sigma)$ is a Gaussian distribution with width (standard deviation) $\tau$ around the mean $\sigma_0$, while $p_3(\sigma)$ is a symmetric double Gaussian distribution with standard deviation $\tau$ and the two maxima at $\sigma_1$ and $\sigma_2$. Distribution 3 has thus the same total mean $\sigma_0$ as distributions 1 and 2. Note that for illustration purposes and the sake of generality the distributions 2 and 3 also allow negative realizations of the property $\sigma$ (which is in principle possible, e.g., for a projected dipole of a particle, or an effective charge that reverses sign under some conditions).  Insertion into eq.~(\ref{1b}) leads to the density profiles $\rho_i(z)$ for the ideal RCs in the external fields. Due to the simple forms of the distributions the problem can be solved analytically and we find  
\begin{eqnarray}
\label{d1}
\rho_1(z)  &=& \rho_1^0\exp(-\beta A \sigma_0 z), \\  \label{d2}
\rho_2(z) &=& \rho_2^0 \exp\left[\frac{1}{2}\beta A z(-2\sigma_0+\beta A\tau^2 z)\right] \;\;{\rm and} \\ \label{d3}
 \rho_3(z) &=& \rho_3^0 \sum_{i=1,2} \exp\left[\frac{1}{2}\beta A z(-2\sigma_i+\beta A\tau^2 z)\right], 
\end{eqnarray}
where the $\rho_i^0$ are constant prefactors. We plot the results for the one-body density profiles $\rho_i(z)$, normalized such that the number of particles per area in all three systems is the same, for selected parameters in Fig.~\ref{fig:id}(a). We choose $\beta A=1$, the size unit $\sigma_0 =1$, $L=15$, and $\tau=1/\sqrt{8}$ for distribution $p_2$. Compared to the non-responsive ideal gas reference, $\rho_1(z)$,  the RC density profiles have lower values at the 'bottom' ($z=0$) and decay less rapidly to zero for large $z$-values. Thus, the system minimizes its free energy by distributing more particles with a smaller property $\sigma<\sigma_0$ at larger $z$.  The density profile $\rho_2(z)$ interestingly exhibits a minimum at $z = \sigma_0/(\beta A \tau^2)=8$ and even rises again for $z > 8$. The reason is that for large $z$ RCs with negative property values are very favorable to be in the high potential regions of the external field. (Increasing $L$ would shift the distribution thus further to negative property values and subsequently to higher densities at large $z$.) In more specific applications there are of course constraints on the property depending on what it represents, e.g., a particle size naturally can not assume negative values.  One finds from application of eq.~(\ref{sigr}) that the position-dependent mean property for distribution 2 is linear in $z$ according to
\begin{eqnarray}
\sigma_2(z) = \sigma_0 - \beta A \tau^2 z,
\end{eqnarray}
being indeed negative for $z>\sigma_0/(\beta A \tau^2)$, see also the dotted lines in Fig.~1(a). We find for the resulting property distributions $N_i(\sigma)$ (per area) the exact result
\begin{eqnarray}
\label{d1}
N_i(\sigma) &=& \rho_i^0p_i(\sigma)\left[\frac{1-\exp(-\beta A \sigma L)}{\beta A\sigma}\right]
\end{eqnarray}
which is plotted for case 2 in Fig.~1(b) together with the unperturbed, single-particle distribution $p_2(\sigma)$. We see clearly the resulting distribution $N(\sigma)$ is shifted to smaller $\sigma$ values, featuring a local maximum in the probability for negative $\sigma$ and a minimum at $\sigma=0$, i.e., a relatively complex, bimodal behavior of $N_2(\sigma)$. 

 For the double-Gaussian distribution $p_3$  we chose  parameter values $\beta A=1$, $\sigma_1 =0.5$, $\sigma_1 =1.5$, $L=15$, and $\tau = 1/\sqrt{32}$. Hence, input is a symmetric double peak with smaller widths for the individual Gaussians than in distribution 2 but with the same mean property $\sigma_0 = (\sigma_1+\sigma_2)/2 =1$. We see in Fig.~1(a) that the qualitative effect for the resulting density profiles is the same as for $\rho_2$ but the rise of density for larger $z$ is much more pronounced, as already half of the original distribution has small property values. Due to the exponential Boltzmann factors the Gaussian peak of the smaller properties dominates more the outcome of the density distributions in the external field than the Gaussian peak of the larger properties. The spatial property distribution 
\begin{equation}
 \sigma_3(z) = \frac{\sum_{i=1,2} e^{\frac{1}{2}A\beta z\left(-2\sigma_i + A\beta\tau^2 z\right) } \left(\sigma_i - A\beta\tau^2 z\right)  }{\sum_{i=1,2} e^{\frac{1}{2}A\beta z\left(-2\sigma_i + A\beta \tau^2z\right)}}
\end{equation}
now features a nonlinear behavior with a crossover between two asymptotic decays at intermediate distances $z \simeq 3$, see Fig.~1(a). This interesting behavior of the mean property in space may be relevant in future applications involving suspensions of RCs having more complex energy landscapes (such as proteins,~\cite{protein,RC} e.g., where slowly vanishing tails or unlikely metastable states of the single-particle property distribution may lead to significant contributions to the liquid structure when subjected to an external field and/or crowded environments.\cite{crowding} 

\subsubsection{Two-body distribution functions and structure-thermodynamics relationships}
\label{2body}
The two-body PDDF of RCs is defined as
\begin{eqnarray}
\rho^{(2)}(\mathbf r,\mathbf r', \sigma, \sigma') = \left\langle \sum_{i=1}^N \sum_{j=1}^N \delta(\mathbf r - \mathbf r_i)\delta(\mathbf r' - \mathbf r_j)\delta(\sigma-\sigma_i)\delta(\sigma'-\sigma_j)\right\rangle,
\end{eqnarray}
which expresses the conditional probability to find a particle with property $\sigma'$ at $\mr'$ if the another one with property $\sigma$ is located at $\mr$.  A normalized pair distribution function can be defined as in standard liquid state theory, via
\begin{eqnarray}
g(\mathbf r,\mathbf r'; \sigma, \sigma') = \rho^{(2)}(\mathbf r,\mathbf r'; \sigma, \sigma')/\rho(\mr,\sigma)\rho(\mr',\sigma')
\end{eqnarray}
which can be shown goes to unity for the ideal gas in the grand canonical ensemble without external fields, where $\rho(\mr,\sigma)\rho(\mr',\sigma') = \rho^2 p(\sigma)p(\sigma')$.  If the system is homogenous and isotropic the pair potential and pair distribution are functions of the distance, $r=|\mr_i - \mr_j|$, only, and we can define a radial distribution function $g(r;\sigma,\sigma')$. For a fixed particle at the origin (test-particle limit) with fixed property $\sigma'$ we obtain the radial one-body PDDF of the surrounding RCs as a function of $g(r;\sigma,\sigma')$, via 
\begin{eqnarray}
\rho(r;\sigma)|_{\sigma'} = \rho N(\sigma)g(r; \sigma, \sigma').
\end{eqnarray}

In the low-density limit we find analogously to non-responsive systems the relation of $g(r;\sigma,\sigma')$ to the pair potential
\begin{eqnarray}
\lim_{\rho\rightarrow 0} g(r;\sigma,\sigma') = \exp[-\beta \phi(r;\sigma,\sigma')] 
\end{eqnarray}
and the one-body PDDF around a single (fixed in the center) RC of property $\sigma'$ in the low-density limit is given by 
\begin{eqnarray}
\lim_{\rho\rightarrow 0} \rho(r;\sigma)_{\sigma'} = \rho p(\sigma)\exp[-\beta \phi(r;\sigma,\sigma')]. 
\end{eqnarray}
Integrating over all $\sigma'$ and $\sigma$ realizations yields the one-body radial density profile around a test-particle in the low-density limit 
\begin{eqnarray}
\lim_{\rho\rightarrow 0} \rho(r) &=& \lim_{\rho\rightarrow 0} \int {\rm d}{\sigma'}p(\sigma')\rho(r; \sigma)_{\sigma'} \nonumber \\
&=& \rho \int {\rm d}{\sigma'}\int {\rm d}{\sigma} p(\sigma')p(\sigma)\exp[-\beta \phi(r;\sigma,\sigma')].
\label{eq:LDL}
\end{eqnarray}

Typically in a coarse-graining procedure, polydispersity of  a property is not accounted for and included in the conventional, effective pair potential $v(r)$. We can map the RC pair potential to the conventional pair potential by comparing eq.~(\ref{eq:LDL}) to the low-density limit for the pair distribution of non-responsive systems, where
\begin{eqnarray}
\lim_{\rho\rightarrow 0} \rho(r) = \rho \exp[-\beta v(r)].
\label{eq:LDLconv} 
\end{eqnarray}
Comparing eqs.(\ref{eq:LDL}) and (\ref{eq:LDLconv}) we find the relation 
\begin{eqnarray}
v(r) = -k_BT \ln \left \{\int {\rm d}{\sigma'}\int {\rm d}{\sigma} p(\sigma')p(\sigma)\exp[-\beta \phi(r;\sigma,\sigma')]\right\}, 
\label{mapping}
\end{eqnarray}
which, not surprisingly, results from integrating out the properties in the pair potental $\phi$.  Considering and respecting relation (\ref{mapping}) in property-resolved liquid models, for example in Monte-Carlo simulations,~\cite{denton:JCP2014, denton:JCP2016,denton:softmatter2015,denton:softmatter2016} is important to avoid double counting of contributions from microstates in the coarse-grained interaction Hamiltonian, where energetic contributions to the property and pair potentials need to be consistently separated.

From (\ref{mapping}) we find the relation between the pair forces ${\rm d}v/{\rm d}r = v' = \langle \phi' \rangle_{\sigma,\sigma'}$, where the subscripts denote an ensemble average of $\phi$ over all property realizations. Hence, the conventional pair potential treatment and the RC framework are consistent in terms of the mean forces in the low density limit. The virial equation for a system of RCs can be derived starting from the pressure expressed by the mean of the internal virial,~\cite{HansenMcDonaldbook} as
\begin{eqnarray}
\beta P &=& \frac{N}{V} - \frac{\beta}{6V} \left\langle \sum_{i\neq j}^N r_{ij}\phi'(r_{ij}) \right\rangle \\
&=&  \frac{N}{V} -  \frac{\beta}{6V^2}\int {\rm d}\mr \int {\rm d}\sigma \int {\rm d}\sigma' N(\sigma)N(\sigma')g(r;\sigma,\sigma') r\phi'(r; \sigma,\sigma').
\label{eq:pressure}
\end{eqnarray}
For low densities we can Taylor-expand the pressure with respect to density ($\beta P = N/V + B^{\rm RC}_2N^2/V^2 + ..$) which defines the second virial coefficient for RCs similar as known for conventional polydisperse systems,\cite{bartlett} as
\begin{eqnarray}
B^{\rm RC}_2 & =&  -\frac{1}{2}\int {\rm d}\mr \int {\rm d}\sigma \int {\rm d}\sigma' p(\sigma)p(\sigma') \{\exp[-\beta \phi(r;\sigma,\sigma')] -1\} \\
&=&  \int {\rm d}\sigma \int {\rm d}\sigma' p(\sigma)p(\sigma') B_2(\sigma,\sigma')
\end{eqnarray}
where $B_2(\sigma,\sigma') = -(1/2)\int{\rm d}\mr \{\exp[-\beta\phi(r;\sigma,\sigma')]-1\}$ is the standard second virial coefficient~\cite{HansenMcDonaldbook} for a pair potential between two particles of properties $\sigma$ and $\sigma'$.
We finally note that the mean potential energy of a bulk system of RCs can be also expressed in terms of their radial distribution function via
\begin{eqnarray}
U &=& \left\langle \sum_i^N \psi(\sigma_i) + \frac{1}{2} \sum_{i\neq j}^N \phi(\mr_i,\mr_j; \sigma_i,\sigma_j)  \right\rangle \nonumber \\
&=&  \int {\rm d}\sigma \psi(\sigma)N(\sigma)+ \frac{1}{2V}\int {\rm d}\mr \int {\rm d}\sigma \int {\rm d}\sigma' N(\sigma)N(\sigma')g(r;\sigma,\sigma') \phi(r; \sigma,\sigma'),
\label{eq:energy}
\end{eqnarray}
where the first term is the potential energy originating from the internal property changes. It can be compared and referenced to the single-particle contribution $U_{\rm id} = \rho V \int {\rm d}\sigma \psi(\sigma)p(\sigma)$. The second term in eq.~(\ref{eq:energy}) is the contribution from the pair interactions. 

%More structure-thermodynamics relations as known for conventional systems, such as for the pressure and compressibility can be derived, but these tasks will be postponed to future work when we aim for a deeper discussion of forces and dynamics of these systems. 

\section{Simple Applications}
\subsection{LDA-DFT for the one-particle density for interacting systems}
\label{LDA}
%
%\begin{figure*}[ht!]
%   \centering
%      \includegraphics[width=0.6\textwidth]{Fig2.pdf}
%   \caption{(a) and (c) one-body density profiles for the ideal property distribution $p_1 = \delta(\sigma-\sigma_0)$ of monodisperse systems for repulsive and attractive $B_2$, respectively, cf. eq.~(\ref{B2}). (b) and (d) The stiff property distribution does not change.}
%   \label{fig2}
%\end{figure*}

We now illustrate the effects of pair interactions on the density and property distributions of RCs. 
For this we resort to the convenient density functional theory (DFT) formalism, the usefulness of which was demonstrated already for conventional polydisperse systems.~\cite{Barrat1986,IgnacioPRL2000} 
To keep it most transparent we use the arguably simplest theory for inhomogeneous density distributions, the local density approximation (LDA) in a second virial expansion.~\cite{HansenMcDonaldbook}  The LDA assumes weak inhomogeneities in the density, $\rho(\mr)/\nabla \rho(\mr) \gg \xi$, that is, density inhomogeneities decay much slower in space than the typical liquid correlation length $\xi$.  In this case we can assume that the system locally obeys an equation of state with a corresponding free energy per volume of the homogeneous fluid $\fex(\rho) = F_{\rm ex}(\rho)/V$.   The grand potential energy within the LDA for RCs can then be written as~\cite{matthias} 
\begin{eqnarray}
\Omega[\rho] &=& k_BT \int {\rm d}\mr \int {\rm d}\sigma \rho(\mathbf r,\sigma)[\ln(\rho(\mathbf r,\sigma)\Lambda^3/p_0) -1] \nonumber \\ &+& \int {\rm d}\mr \,\fex(\rho(\mathbf r,\sigma)) + \int {\rm d}{\mr}\int {\rm d}\sigma\, \rho(\mathbf r,\sigma) [\phi_{\rm ext}(\mathbf r,\sigma)+\psi(\sigma)-\mu],
\end{eqnarray}
where the first term on the right hand side is the ideal gas free energy, the second is the excess free energy $F_{\rm ex}$, and the last term couples to the external and chemical potentials. Following our notations above, the local excess free energy per volume on the second virial level reads
\begin{eqnarray}
f_{\rm ex}(\mr) =  k_BT \int {\rm d}\sigma  \rho(\mathbf r,\sigma)  \int {\rm d}\sigma' \rho(\mathbf r,\sigma')B_2(\sigma,\sigma'), 
\end{eqnarray}
 where we used the LDA approximation $\rho(\mr, \sigma)\simeq \rho(\mr',\sigma)$ and thus avoided a convolution in space.  Minimization ($\partial \Omega/\partial \rho=0$) leads to
\begin{eqnarray}
\rho(\mathbf r,\sigma)= q\exp[-2\int {\rm d}\sigma' \rho(\mathbf r,\sigma')B_2(\sigma,\sigma')  -\beta \psi(\sigma) -\beta  \phi_{\rm ext}(\mathbf r,\sigma))]. 
\label{eq:rho}
\end{eqnarray}
which is the final result in the LDA in the second virial limit. Note that it still involves a convolution over the property $\sigma$. We can further simplify by assuming that the property distribution $N(\sigma)$ is relatively narrow and unimodal and can be approximated by a $\delta$-function peaking at $\sigma_0 \simeq \sigma$, resulting in $\int {\rm d}\sigma' \rho(\mr,\sigma')B_2(\sigma,\sigma') \simeq  \rho(\mr)B_2(\sigma)$. With that, eq.~(\ref{eq:rho}) reduces to
\begin{eqnarray}
\rho(\mathbf r,\sigma)= q\exp[-2\rho(\mathbf r) B_2(\sigma)  -\beta \psi(\sigma) -\beta  \phi_{\rm ext}(\mathbf r,\sigma))]. 
\label{eq:rho2}
\end{eqnarray}
In this form the conventional one-body density $\rho(\mr)$ appears and avoids the $\sigma$-convolution, and it is particularly simple to recognize the effects of the non-vanishing pair interactions on the local structure and properties. Locally high densities will be modified by the $B_2$-term which is not only a function of space but also of property.  For an illustration of these contribution to the distributions in the presence of an external field we chose the example in Fig.~1, that is the linear field $\phi_{\rm ext}(z,\sigma) = A\sigma z$ and the distributions $p_1(\sigma)$ and $p_2(\sigma)$ according to eqs.~(\ref{p1}) to (\ref{p2}), respectively, and solve the non-linear eq.~(\ref{eq:rho2}) numerically. (For the bimodal distribution $p_3(\sigma)$ in eq.~(\ref{p3}) we should solve eq.~(\ref{eq:rho}) but we refrain to do it because it does not add qualitative insight.)

\begin{figure*}[ht!]
   \centering
      \includegraphics[width=0.8\textwidth]{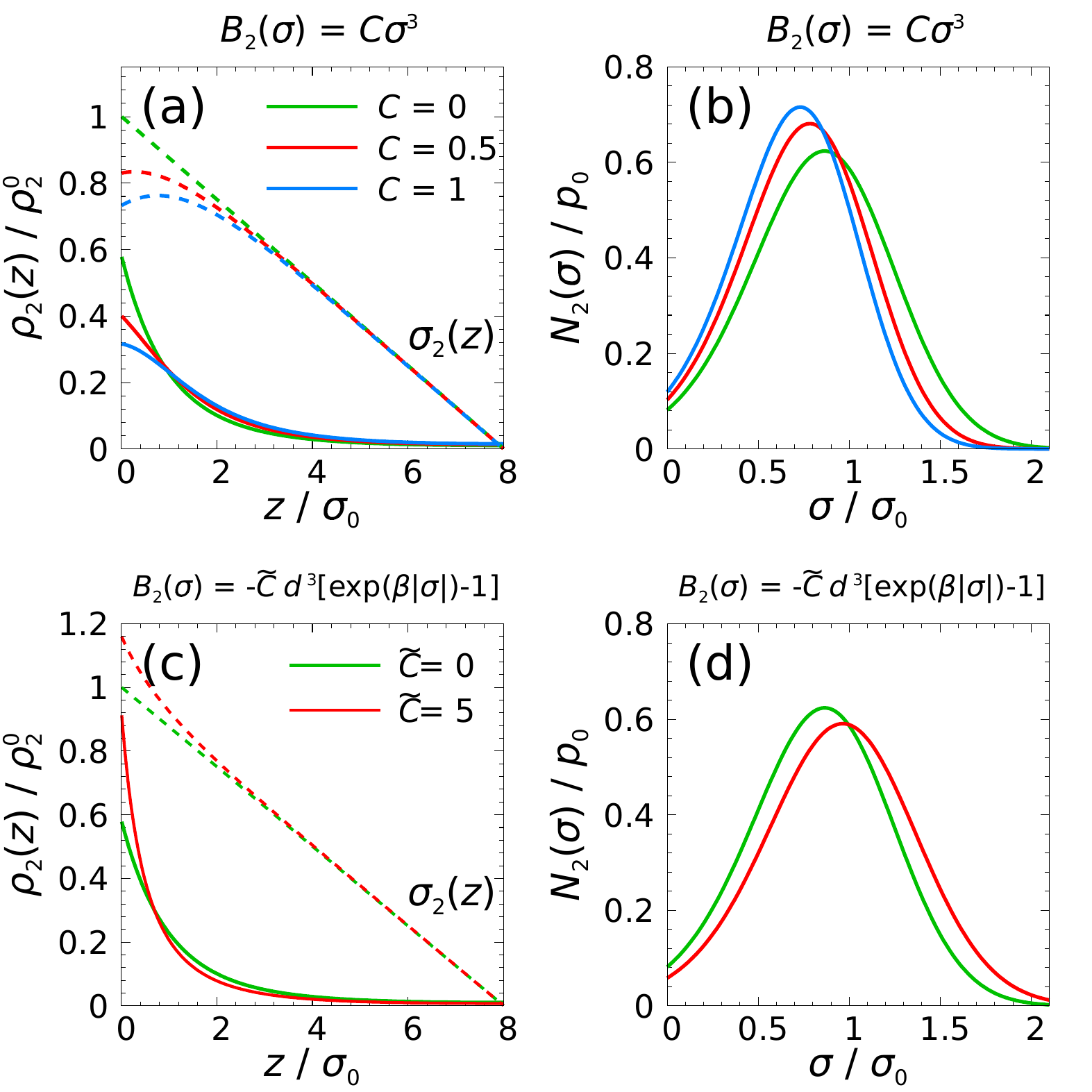}
   \caption{(a) and (c): Number density profiles for the Gaussian ideal property distribution $p_2(\sigma)$, cf. eq.~(\ref{p2}), of monodisperse systems for repulsive and attractive $B_2$, cf. eq.~(\ref{B2}), respectively. (b) and (d) The corresponding emerging property distributions, $N(\sigma)$, in the interacting systems depend on the interactions: the more repulsive (attractive) the interactions, the smaller (larger) the mean property $\sigma$.}
   \label{fig3}
\end{figure*}

The functional form of $B_2(\sigma)$ depends on the specific property that $\sigma$ represents. For instance, in the case of simple excluded-volume interactions $B_2\propto \sigma^3$, where in this case $\sigma$ is the particle excluded size. However, if $\sigma$ is an electrostatic surface potential or an attraction energy that serves as a prefactor in a pair potential, then $B_2$ typically depends exponentially on such a property. For example, for a square well potential with attractive depth $\sigma$, $B_2 \propto [\exp(\beta|\sigma|)-1]$. We thus chose the following cases in our parametric study: 
\begin{eqnarray}
{\rm I:} \;\;\; B_2(\sigma) &=& C\sigma^3  \;\;\;{\rm and} \\
{\rm II:} \;\;\; B_2(\sigma) &=& - \tilde C d^3 [\exp(\beta|\sigma|)-1] 
\label{B2}
\end{eqnarray}
The parameters $C>0$ and $\tilde C>0$ are dimensionless prefactors that define the magnitude of the $B_2$-coefficient. We introduced a particle size $d$ for case II to properly define a particle size and packing fraction. (In case I this is not necessary as $\sigma$ represents already the particle size.) We consider moderate densities  in the range of $\int {\rm d}z \rho(z) \simeq  L/\sigma^3$ or $ \simeq L/d^3$ for cases I and II, respectively, to observe significant interaction effects.  The numerical results for $\rho(z)$, $N(\sigma)$, and $\sigma(z)$ for the Gaussian distribution $p_2$ are shown in Fig.~2.  The density profiles displayed in Fig.~2(a) and (c) show the expected $B_2$ effects, i.e., decreasing/increasing density for increasingly repulsive/attractive interactions, respectively. Importantly, however, the interactions change the local property $\sigma(z)$ significantly at high densities (small $z$) as well as the overall distribution $N(\sigma)$, shown in panels (b) and (d) of Fig.~2. For repulsive interactions, the distribution shifts to the left, favoring smaller properties, and opposite for attractive interactions.  

A better treatment beyond LDA must include more sophisticated nonlocal excess functionals~\cite{HansenMcDonaldbook,IgnacioPRL2000,matthias} which shall be addressed in future work. We expect very interesting results in particular for very broad or even polymodal property distributions in dense, crowded systems with strong spatial correlations which are not well approximated in the simple LDA framework introduced above. 

\subsection{Structure of hard spheres with highly localized sizes in a planar slab confinement}

An interesting and so far not much discussed fact for conventional polydisperse as well as RC systems is that the action of particle interactions, expressed by the pair potential $\phi(r;\sigma,\sigma')$, can be quite localized under the action of an external field. The latter imposes local properties, or can even select properties from the ideal distribution, which has large implications on the local number density.  We now have a closer look at the explicit consequences of highly localized properties on the density distribution.  For this, we consider a system of RCs confined between two hard walls separated by a distance $L_z$ in the $z$-direction. The property $\sigma$ is now the particle size. The system is homogeneous in lateral dimensions $x$ and $y$, and we look for the density profile $\rho(z)$. With a conventional (i.e., constant in space) hard sphere pair potential 
\begin{eqnarray}
V_{\rm HS}(r) = \begin{cases}
\infty & r \leq \sigma_0 \\
0 & \, \text{else}
\end{cases}
\end{eqnarray}
this problem was considered and solved many times in various approximations.\cite{HansenMcDonaldbook} 
Now, we impose a position-dependence of the pair potential through highly localized properties 
\begin{eqnarray}
V_{\rm HS}(|\mathbf r_2 - \mathbf r_1|;s(z)) = \begin{cases}
\infty & |\mathbf r_2 - \mathbf r_1| \leq [{\sigma(z_1)+\sigma(z_2)}]/{2} \\
0 & \, \text{else}, 
\end{cases}
\end{eqnarray}
using a linear scaling of the hard core size with a spatially-dependent function $s(z)$ as
\begin{eqnarray}
\sigma(z)  = \sigma_0 s(z).   
\end{eqnarray}
We further assume that the interaction size of a RC changes along the $z$-axis with the linear relation
\begin{equation}
 s(z) = \left[ 1 + fz/L_z\right],
 \label{eq:s}
\end{equation}
where $f$ is a scaling factor, so that $\sigma(0) = \sigma_0$ and $\sigma(L_z) = (1+f)\sigma_0$. We will study two scaling factors $f=0.2$ and $f=-0.8$. An artistic illustration of such a size-inhomogeneous system for $f=-0.8$ is presented in Fig.~\ref{fig3}. The scaling function $s(z)$ can be generated by an external  potential of the form 
\begin{eqnarray}
-k_BT\ln \phi_{\rm ext}(z,\sigma) = \delta (\sigma - \sigma_0 s(z)).
\label{eq:psidelta}
\end{eqnarray}
Due to this extreme external constraint, the ideal distribution $p(\sigma)$ does not play a role anymore; the external potential defines strictly the value of $\sigma$ of the RC at position $z$. As a consequence we obtain a  modified form of the local hard sphere packing fraction through
\begin{eqnarray}
\eta(z) = \frac{\pi}{6}\rho(z)\sigma(z)^3  = \frac{\pi}{6}\rho(z)\sigma_0^3s(z)^3,   
\end{eqnarray}
where the sphere volume is now a position-dependent quantity. 
\begin{figure}[ht!]
   \centering
      \includegraphics[width=0.8\textwidth]{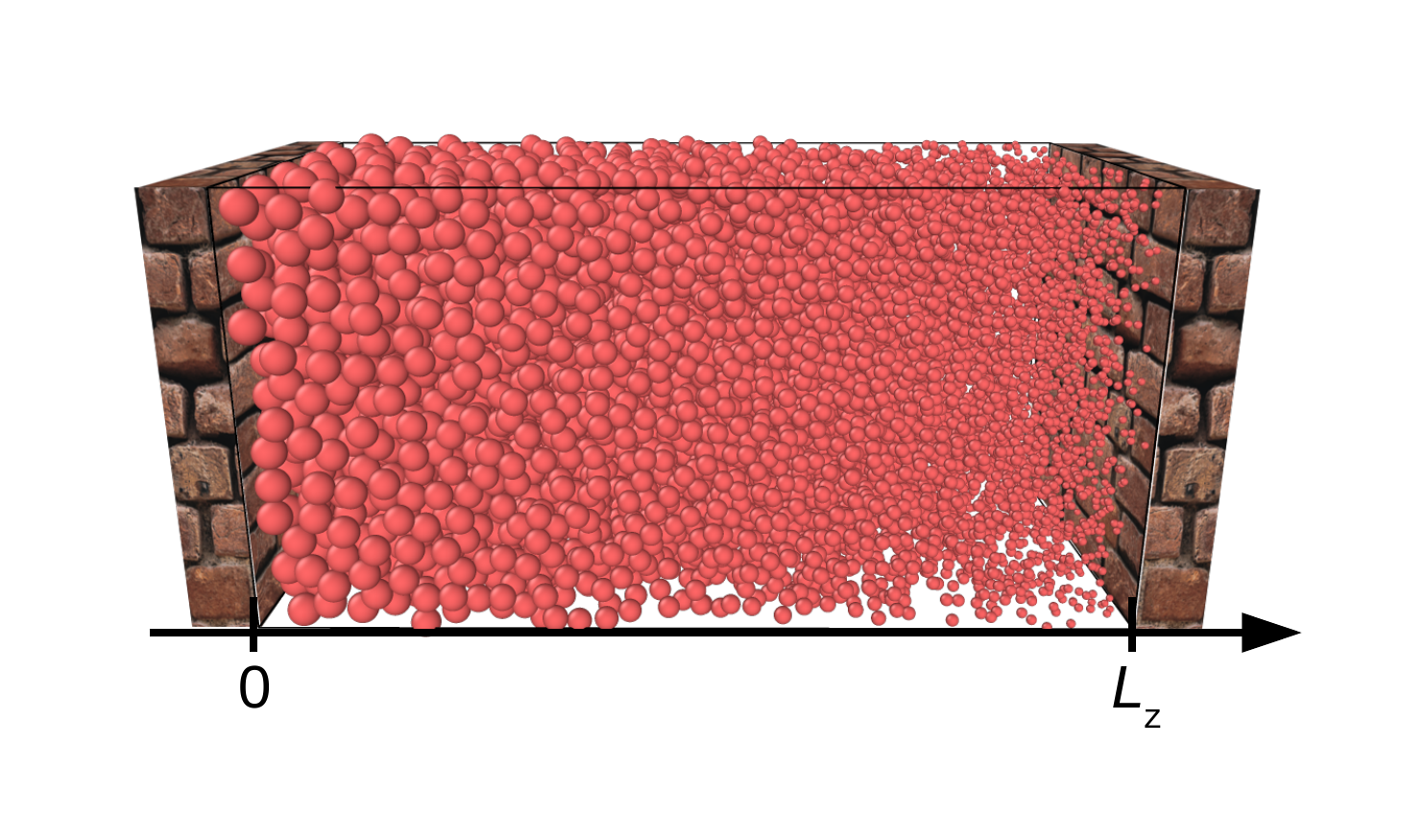}
   \caption{Artistic illustration of hard spherical RCs with highly localized properties (sizes) confined between two hard walls in distance $L_z=25\sigma_0$ in $z$-direction. Sphere diameters decrease linearly from $\sigma(z=0)=\sigma_0$ to $\sigma(z=L_z)=0.2\sigma_0$ (scaling factor $f=-0.8$ in the linear scaling function eq.~(\ref{eq:s})).}
   \label{fig4}
\end{figure}

For further simplification, let us assume $s(z)$ is a very slowly varying function, i.e., $s/(\partial s/\partial z) \gg \sigma_0$ so that we can work in the LDA framework introduced above. Under this assumption, we can approximate the interaction diameter as $[{\sigma(z_1)+\sigma(z_2)}]/{2} =\sigma_0[{s(z_1)+s(z_2)}]/{2} \simeq \sigma(z)$, where $z = (z_1+z_2)/2$. The local interaction then simplifies to 
\begin{eqnarray}
V_{\rm HS}(r;s) = \begin{cases}
\infty & r \leq \sigma(z) \\
0 & \, \text{else}
\end{cases}
\end{eqnarray}

We make use of the LDA equation (\ref{eq:rho}) as a function of $z$ which, after inserting the external potential (\ref{eq:psidelta}) reduces to 
\begin{eqnarray}
\rho(z,\sigma)= q\exp[-\beta \fex'(\rho(z,\sigma))]_{(\sigma\equiv\sigma(z))}, 
\label{HSLDA}
\end{eqnarray}
with a prescribed position-dependent $\sigma(z)$. In the case of hard spheres, it is possible to go beyond the $B_2$-level, since a reasonably accurate approximation of $\fex$ is provided by the Carnahan-Starling expression:~\cite{HansenMcDonaldbook} 
\begin{eqnarray}
\beta \fex(\rho) = \rho \frac{4\eta- 3\eta^2}{(1-\eta)^2}. 
\label{eq:CS}
\end{eqnarray}

\begin{figure*}[ht!]
   \centering
      \includegraphics[width=0.8\textwidth]{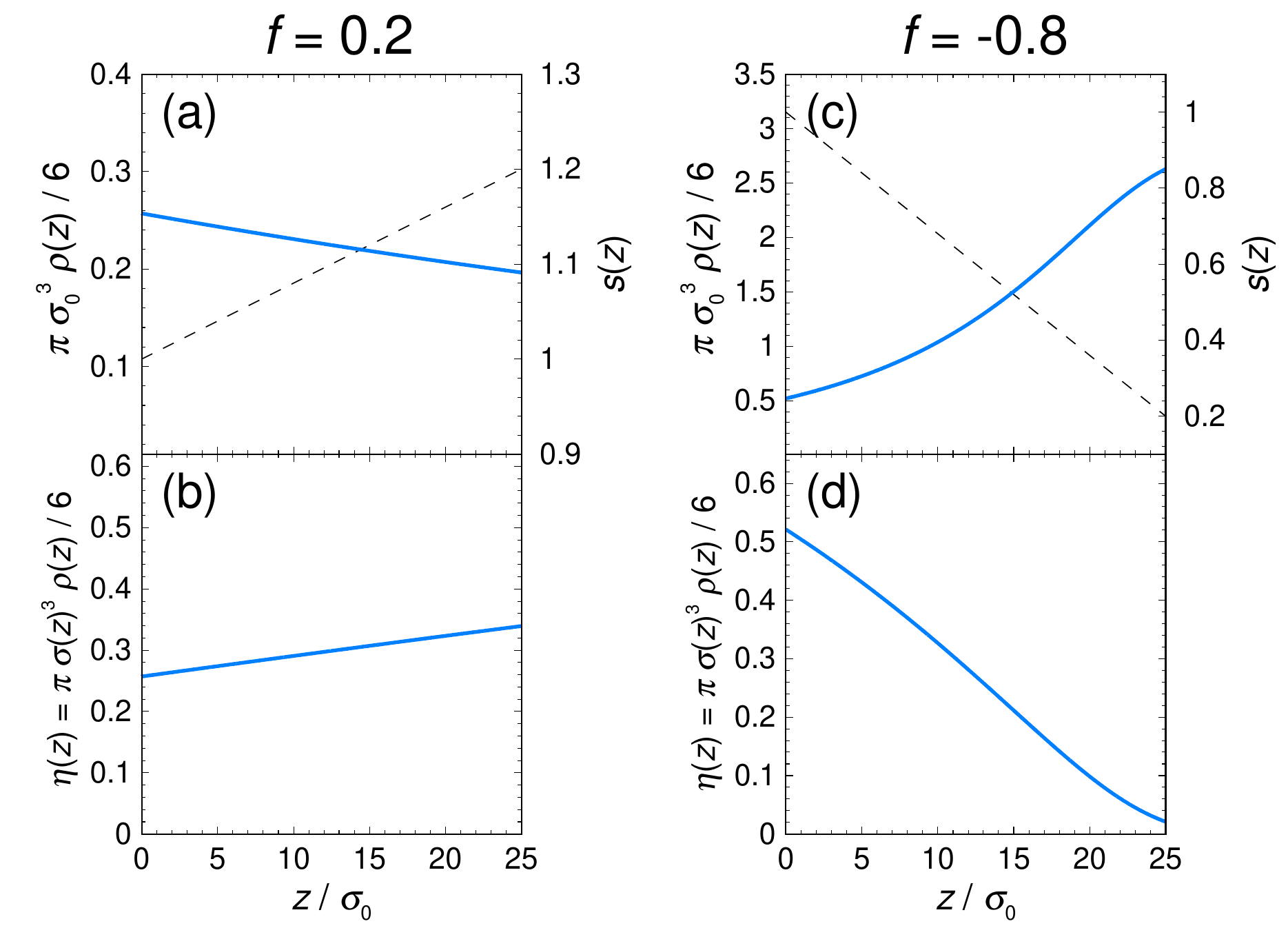}
   \caption{Scaled density profiles, $\rho(z)\sigma_0^3\pi/6$, and packing fraction $\eta(z)=\rho(z)\sigma(z)^3\pi/6$ of hard spherical RCs confined between two walls, cf.~Fig.~3, separated by a distance $L_z=25\sigma_0$ from solution of the LDA equation (\ref{HSLDA}).  (a) and (b) are for a scaling factor 0.2, i.e., sphere diameters increase from 1 to 1.2$\sigma_0$ from left to right (see dashed line and right scale in (a)). (c) and (d) are for a scaling factor $f=-0.8$, i.e., sphere diameters decrease from 1 to 0.2$\sigma_0$ from left to right (see dashed line and right scale in (c)).}
   \label{fig5}
\end{figure*}

We solve eq.~(\ref{eq:CS}) numerically for a fixed density per area $\int_0^{L_z}\rho(z){\rm d}z = 10.8/\sigma_0^2$ and $66.3/\sigma_0^2$ for the scaling factors $f=0.2$ and $f=-0.8$, respectively. The resulting density and packing fraction profiles are presented in Fig.~\ref{fig5} for the two examples, $f=0.2$ in (a) and (b), and $f=-0.8$ in (c) and (d), respectively. The scaling behavior $s(z)$ is depicted by dashed lines in (a) and (c) (right vertical axes). Note that the number density profiles are scaled by $\sigma_0^3$ while the packing fraction $\eta(z)$ is determined by scaling $\rho(z)$ with $\sigma(z)^3$. The resulting profiles are far from trivial. For the weak scaling $f=0.2$ we observe an almost linear decrease of density with increasing diameter along the $z$-direction, while the packing $\eta(z)$ linearly increases. The qualitative behavior of the density profile can be readily understood by the necessity of maintaining constant pressure in equilibrium. Larger colloids have higher osmotic pressure at the same number density, so that their local density needs to decrease to maintain mechanical equilibrium. This effect does not cancel exactly with the effects of the local volume change and as a net result the packing fraction shows the opposite trend and increases with growing sphere size.  For the larger scaling factor, $f=-0.8$, the effects are qualitatively the same but more drastic. The number density increases now in a highly nonlinear fashion with decreasing RC size, by about a factor of 5, while the packing fraction close to the right wall almost vanishes. 

Even though more sophisticated (non-local) DFT approaches~\cite{HansenMcDonaldbook,IgnacioPRL2000,matthias} need to be employed to capture better any stronger inhomogeneites of the profiles, e.g., spatial correlations (layering) at the wall for high local packing fractions,  this simple example already demonstrates the rich and non-trivial behavior of RCs with highly localized properties, which can be captured by the present approach. 

\subsection{Environmental action of an additional cosolute component}

In many cases relevant for applications, a stimulus on the RCs will be imposed not by an external field but by additional cosolute components, or 'crowders', changing locally the properties of the considered RCs by direct interactions. Then the application of a one-component theory as above is limited because the RCs couple back to the cosolute environment. In addition, application of mechanical balance (constant osmotic pressure) arguments must include all degrees of freedom in the system, including those of the cosolute environment.  

In order to show how the present theoretical developments can deal with such situations, we illustrate on a specific, yet practically relevant, model system. We consider a system consisting of two solutes immersed in an implicit solvent.
The first solute is non-responsive (this could be a salt or an osmolyte, see, e.g., the literature in a previous work~\cite{Heyda2014a}, or crowding macromolecules\cite{crowding}), while the second are RCs, whose properties depend on the local number density of the first solute. As a result, it is necessary to resolve the distribution of both species and not only of the RCs. Inspired by the Asakura-Oosawa model for depletion interactions~\cite{HansenMcDonaldbook} to account for soft deformations and compression of polydisperse polymers in colloid-polymer mixtures,~\cite{matthias} we further assume for simplicity that species 1 does not interact with itself.
Within the LDA-DFT framework on the simplest second virial level (cf. eq.~(\ref{eq:rho2}) in section~\ref{LDA}) , the grand potential functional of such a system can be written
 \begin{eqnarray}
\Omega[\rho_1,\rho_2] &=& k_BT \int {\rm d}^3\,r \rho_1(\mathbf r)[\ln(\rho_1(\mathbf r)\Lambda_1^3) -1] \nonumber
+  k_BT \int {\rm d}\mr \int {\rm d}\sigma \rho_2(\mathbf r,\sigma)[\ln(\rho_2(\mathbf r,\sigma)\Lambda_2^3/p_0) -1] \\ 
&+&  k_BT \int {\rm d}\mr\int {\rm d}\sigma \; 2B_2^{12}(\sigma)\rho_1(\mathbf r)\rho_2(\mathbf r,\sigma) \nonumber + \int {\rm d}\mr\int {\rm d}\sigma\; B_2^{22}(\sigma)\rho_2(\mathbf r,\sigma)^2\\ 
&+& \int {\rm d}^3\,r \rho_1(\mathbf r) [\phi_{{\rm ext},1}(\mathbf r)-\mu_1]+ \int {\rm d}{\mr}\int {\rm d}\sigma\, \rho_2(\mathbf r,\sigma) [\psi((\sigma))-\mu(\sigma)] \nonumber
\end{eqnarray}
The first two terms on the right-hand-side represent the ideal gas contributions of the two species $i=1,2$. In the third and fourth terms, the $B_2^{ij}(\sigma)$ describe the interaction between $i$ and $j$ in terms of the standard second virial coefficient. The last two terms describe the one-body terms due to external fields and chemical potentials, where we have further assumed that the external field acts directly only on species 1 (but, as shown below, this still leads to an indirect effect on the RCs). Minimization ($\delta \Omega/\delta \rho_i = 0$) yields the coupled equations
 \begin{eqnarray}
\rho_1(\mathbf r) &=& q_{1} \exp[-2\rho_2(\mathbf r)B_2^{12}(\sigma) - \beta \phi_{{\rm ext},1}(\mathbf r)] \nonumber \;\;\; {\rm and} \\
\rho_2(\mr,\sigma) &=& q_2\exp[-2\rho_1(\mr) B_2^{12}(\sigma)-2\rho_2(\mr)B_2^{22}(\sigma) -\beta \psi(\sigma)] 
\label{LDAcosolute}
\end{eqnarray}
where the $\rho_i(\mr)$ (without $\sigma$ argument) are all one-body number densities.  (We consistently used the approximation of a $\delta$-like property distribution, such that $\int {\rm d}\sigma 2B_2^{12}(\sigma)\rho_2(\mr,\sigma) \simeq 2 B_2^{12}(\sigma)\rho_2(\mr)$.) 
Without solving explicitly these equations, it can be seen that species 2 distributes inhomogeneously in space (in both translation and property), because the inhomogeneous spatial distribution of species 1 propagates to species 2 via the interactions, expressed by the coupling with the $B_{12}$ terms.  In other words, the osmotic pressure from species 1, which is not constant in space due to the presence of the external field, has to be balanced by an inhomogeneous space and property distribution of species 2. The local properties of RCs affect both the $B_{12}$ cross-term and the $\fex$ term. In particular, we recognize that the $\rho_1B_2^{12}$-term acts as a $\sigma$-dependent external potential on the RCs, so that one could define an effective one-component RC system $\phi_{\rm ext}(\mr) = k_BT \rho_1(\mr)B_2^{12}(\sigma)$, if $\rho_1(z)$ were fixed. Note that in the absence of an external field (no inhomogeneity and $\rho_1$ and $\rho_2$ are constant), the property distribution of the RCs changes according to 
 \begin{eqnarray}
N_2(\sigma) &=& q_2V\exp[-2\rho_1B_2^{12}(\sigma) - 2\rho_2 B_2^{22}(\sigma) -\beta \psi(\sigma)]   
\end{eqnarray}
due to the presence of cosolute 1. 
 
The $B_2^{12}(\sigma)\rho_1$ term is actually related to a very common model in the large body of studies of cosolute/salt effects on protein folding/unfolding or the coil-to-globule transitions of polymers. A discussion of this aspect can be found in Ref.~\cite{Heyda2014a}: it was empirically observed that for a wide range of systems the free energy difference between two states (unfolded versus folded protein) is linear in cosolute concentration $c$, i.e., follows the linear law $\Delta \Delta G(c) = \Delta G({\rm folded})(c) - \Delta G({\rm unfolded})(0) = mc$, where the coefficient $m$ is the so-called $m$-value. The physical origin of $m$ is argued to arise from preferential desorption or adsorption of cosolutes to the macromolecule, leading to depletion-induced collapse or attraction-induced swelling of the macromolecule, respectively. If salt plays the role of the cosolute, the community talks about 'salting-out' and 'salting-in' with respect to the macromolecular size or solubility of a dispersion of those macromolecules. The theoretical framework introduced in the present work provides a basis to understand the emergence of such a scaling of the free energy difference between states with the cosolute concentration $c$, since the coupling term $B_2^{12}(\sigma)\rho_1$ can be readily re-expressed in terms of the above $m$-value and $c$.

\begin{figure}[ht!]
   \centering
      \includegraphics[width=0.4\textwidth]{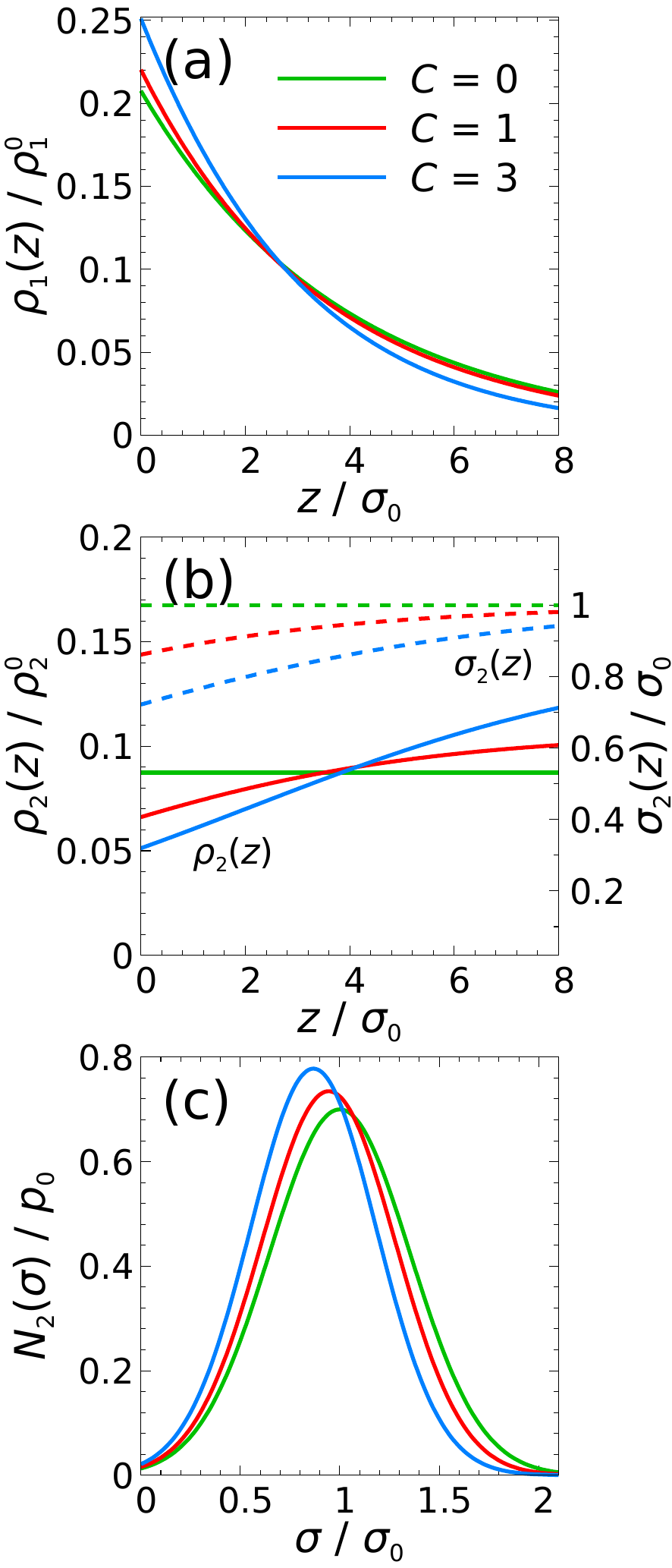}
   \caption{Distributions of a two-component system (cosolutes and RCs ) between two walls separated by a distance $L=8$ according to eq.~(\ref{LDA2}). A linear external field $\phi_{{\rm ext},1}(z) = {\tilde A}z$ acts only on the non-responsive, mutually ideal species 1, while the RCs (species 2) are interacting with species tuned by coupling parameter $C$ (legend), see text. The spatial distribution of cosolute species 1 (a) couples to the density profiles and property distributions of the RCs (species 2) shown in panel (b) and (c), respectively. }
   \label{fig:6}
\end{figure}

The remaining question of how $B_2$ depends on the coarse-grained variable $\sigma$ needs to be inferred for specific cases from a coarse-graining procedure formally along the lines defined in section II.A. As a simple illustration, we consider here a parametric study and use the definitions analogous to eq.~(\ref{B2}), $B_2^{12}(\sigma) = C\sigma^3 $, i.e., $\sigma$ could represent an effective exclusion size between the two species. For simplicity, we assume that also the RCs are ideal among themselves, that is $B_2^{22}=0$. We study only inhomogeneities in the $z$-direction.  With these specifying assumptions, eq.~(\ref{LDAcosolute}) simplifies to
 \begin{eqnarray}
\rho_1(z) &=& \rho_1^0 \exp[-2\rho_2(z)B_2^{12}(\sigma) - \beta \phi_{{\rm ext},1}(z)] \nonumber \;\;\; {\rm and} \\
\rho_2(z,\sigma) &=& \rho_2^0 p_0 \exp[-2\rho_1(z)B_2^{12}(\sigma) -\beta \psi(\sigma)],  
\label{LDA2}
\end{eqnarray}
where we replaced the normalizing prefactors by $\rho_1^0$ and $\rho_2^0p_0$ and we consider an area density $\int_0^L {\rm d}z\rho_i(z) = 0.7/\sigma_0^2$.  Numerical solutions for the linear external field $\phi_{{\rm ext},1}(z) = {\tilde A}z$ between hard walls separated by a distance $L=8$ and are shown in Fig.~5. We chose $\beta \tilde A =
-\ln(0.125) / L$ such that $\rho_1(L) = 0.125\rho_1(0)$ for $C = 0$ and report results for $C=0,1$ and $3$. The density distribution of species 1 plotted in panel (a) shows the expected exponential decay but depends on $C$ due to the coupling to the distribution of species 2, which also distributes inhomogeneously, cf. Fig.~5(b). The repulsion between the cosolutes and RCs (positive $B_{12}$) not only results in a depletion of RCs in the region enriched in cosolute (small $z$), as expected, but also in a significant decrease of the local size of the RCs $\sigma$: the higher osmotic pressure due to the cosolute compresses the RCs. The consequence is an overall shifted distribution of sizes, $N(\sigma)$, to lower values, as displayed in Fig.~5(c). This is consistent with the DFT study of simple soft polymers in bulk compressed by hard-sphere colloids.~\cite{matthias} Here, we offer a different perspective and argue that similar physics applies in general when considering property changes induced by the co-solvent environment, e.g., (de)stabilization of responsive structures, e.g., a protein or thermosensitive polymer by molecular osmolytes~\cite{Heyda2014a} or macromolecular crowders.~\cite{crowding} 

\section{Concluding remarks}

We presented the fundamental statistical mechanics of inhomogeneous colloidal suspensions of responsive colloids (RCs), where we extended the conventional coarse-grained picture of simple colloidal liquids by keeping a generic 'property' degree of freedom in addition to the position of the center-of-mass of a RC.  The collective property variable describes the internal state of the RC and the effect of the latter on the interactions with other colloids. We showed that the distribution of this collective variable (which is defined by a prescribed free energy landscape for an isolated RC) can change under the action of external fields and interactions, and that such a coupling results in a rich behavior in terms of spatial and property distribution of the RCs. In equilibrium and in the thermodynamic limit, the statistical mechanics framework for RCs is equivalent to that for conventional polydisperse systems (for which the property per particle is fixed) but our perspective reveals new aspects and opens the way to promising new research direction on RCs.

For example, a single RC often displays a complex polymodal property distribution, indicating that each particle can fluctuate between distinct multiple states.~\cite{protein,RC, IDP_switching,DNA,chiwu} Polymodal parent distributions have not received much attention yet in the literature. The consequences on structure, response, and phase behavior of fluids of polymodal RCs are expected to be richer than previously studied polydisperse systems and are thus worth exploring further, both in homogeneous bulk and under the action of external fields, providing a handle to manipulate and localize the properties of the RCs. The focus on more complex one-particle property distributions could be in particular interesting for reverse engineering problems, where properties and their energy landscapes of colloids and materials need to be optimized and adapted for desired functionalities. Another fascinating prospect is certainly to investigate the dynamics of these systems, e.g., diffusion, rheology, or internal single particle kinetics, which should be very different from that of conventional polydisperse systems because single RCs respond with property changes on a large spectrum of timescales, or can be even kinetically trapped after some responsive switch. 

Finally, RCs  can be made active, i.e., involving self-fueled, time-dependent property changes (such as a biological cell that actively changes shape\cite{shape3} or active polymers~\cite{oscillating, Heuser, DNA_hydrogel,breathing}), which immediately inspires many new questions on structure and dynamics of these non-equilibrium systems with active property changes and active interactions,\cite{moncho} demonstrated already for a wide range of systems exhibiting active motility.~\cite{RevModPhys88}
 
\begin{acknowledgments}

The authors are indebted to Jean-Pierre Hansen, Upayan Baul, and Arturo Moncho-Jord{\'a} for helpful discussions and a critical reading of parts of the manuscript. This project has received funding from the European Research Council (ERC) under the European Union's Horizon 2020 research and innovation programme (grant agreement Nr. 646659). B.R. acknowledges the financial support from the Alexander von Humboldt foundation via the Bessel research award.

\end{acknowledgments}

%\bibliography{literature}
 
\end{document}